# Gyro-Control of a Solar Sailing Satellite

By Hendrik W. JORDAAN[1)] and Willem H. STEYN[1)]

[1)]*Electric and Electronic Engineering Department, Stellenbosch University, Stellenbosch, South Africa*

Recent successes in the deployment of sails in space have reduced the risk associated with solar sailing missions.   The attitude control requirements for a solar sailing mission is low with only slow attitude maneuvers needed to maintain a stable attitude and produce a required solar thrust.   Future science missions will require large attitude maneuvers with a fully deployed sail.   This article investigates the current options for attitude control on solar sails and proposes a gyro-controlled solar sailing.   This concept uses a spinning solar sail to construct a control moment gyroscope capable to produce large torques.   Steering laws for performing attitude maneuvers and simulation results are presented which demonstrates the capabilities of such a solution.

**Key Words:**   ADCS, solar sailing, gyro-control, spinning sail

**Nomenclature**

| | | |
|---|---|---|
| $q$ | : | quaternion vector |
| $\omega$ | : | angular rate vector |
| $I$ | : | moment of inertia tensor |
| $A$ | : | transformation matrix |
| $h$ | : | angular momentum vector |
| $u$ | : | control signal |
| $K$ | : | gain matrix |

Subscripts

| | | |
|---|---|---|
| $B$ | : | body or satellite bus |
| $O$ | : | orbit frame |
| $S$ | : | sail system |
| $C$ | : | momentum countering system |

## 1. Introduction

Advancements in solar sailing deployment techniques and mechanisms in the recent past and successful demonstrations in space[1-4] have reduced the risk of using solar sails as part of satellite missions.   Currently solar sailing is seen as the main objective of all satellite missions containing solar sails.   The attitude control requirements for such a solar sailing mission is low with only slow attitude maneuvers needed to maintain a stable attitude and produce a required solar thrust.   The attitude requirements might greatly increase in satellite missions containing other payloads than only the solar sail. Many mission payloads require fast and accurate attitude maneuvers for the payload to perform to its specifications. Maneuvers must now be completed successfully with a deployed sail.   Current sailing attitude determination and control systems (ADCS) do not have the capabilities to perform high rate maneuvers.   Fast attitude control will result in the excitation of high frequency oscillations of elements of large deployable structures.   These unwanted oscillations will influence the attitude stability and disturb satellite payloads with high sampling rates or long exposure times.

This article will investigate the attitude performance related to different control methods.   A control moment gyroscope (CMG) controlled tri-spin satellite and the related steering laws is presented as another option to perform high rate maneuvers. The steering laws are implemented in a simulation and the effect of a number of control inaccuracies are investigated.

## 2. Attitude Control Parameters

The largest difference between a solar sailing satellite and a standard satellite is the large moment of inertia (MoI), which is obtained having a large deployed structure.   Even if the mass of the sail remains small, the area of current proposed solar sails will result in a substantial MoI comparable with much larger conventional spacecraft.   This requires the attitude control method to generate large control torques to provide the attitude performance which some space missions require.

### 2.1. Sail/Spacecraft MoI Ratio

Assume a standard four boom square solar sail with an area of 80$m^2$ with a sail density of 4.8$g/m^2$ and the 6.5m booms each with density of 45g/m.   By simply increasing the sail area to 100$m^2$ the MoI of the sail system increases by 43.9%, which correlates to less than a meter increase in each of the four booms, see Table 1.

Table 1.   Moment of inertia increase between 80$m^2$ and 100$m^2$ sails

| Parameter | 80m2 Sail | 100m2 Sail |
|---|---|---|
| Boom length $\ell$ | 6.325m | 7.071m |
| MoI $I_{xx} = I_{zz}$ | 10.149kg.m2 | 14.607kg.m2 |
| MoI $I_{yy}$ | 20.299kg.m2 | 29.213kg.m2 |

This requires the attitude control actuator's specifications to also increase if similar attitude performance is required.   The increase in actuator specifications are determined by the sail/spacecraft moment of inertia ratio, and is hard to obtain without an increase in mass.   An increase in mass will reduce the characteristic acceleration.   The moment of inertia tensor matrix of the sail $I_S$ is element-wise divided by the total spacecraft tensor matrix $I$

$$\Lambda = I_S / I \quad (1.)$$

with the sail/spacecraft moment of inertia ratio, the maximum element within the resulting matrix

$$\lambda = \max(\Lambda) \quad (2.)$$

This ratio highlights a number of aspects concerning the attitude dynamics. A small $\lambda$ indicates that the rotational dynamics of the spacecraft is dominant. A small change in the MoI in the sail will not have a large effect on the spacecraft. This is ideal but might indicate a low characteristic acceleration of the solar sailing spacecraft. A larger $\lambda$ indicates that the dynamics are greatly influenced by the sail, thus any changes in MoI will greatly influence the attitude performance whether this is a change in sail area or vibrations of the non-rigid elements of the sail.

### 2.2. Current Attitude Control Methods

A number of different attitude control mechanisms of solar sails exist. Conventional attitude control is dependent on reaction wheels, control-moment gyroscopes (CMG), or small thrusters normally mounted on the corners of the satellite.

Solar sails generates a solar thrust, which can be modelled as a certain resultant force at the center-of-pressure (CoP). A number of solar sail specific attitude control methods use the aspect of changing the vector between the center-of-mass (CoM) of the spacecraft and the CoP. A CoM-to-CoP offset will create the required torque for the satellite to change its orientation. IKAROS changed the reflective properties of certain sections of the sail to move the CoP[1]. This method can be used to generate control torques in 2 axes. Other conceptual solar sails use control vanes[5]. These are sections of reflective material which angle relative to the sun can be changed. If these control vanes have one degree of freedom (DOF) then 2 axes can be controlled, but if 2-DOF rotations are possible then all 3 axes of the spacecraft can be controlled. Control vanes also use the concept of changing the generated solar thrust vector to induce a control torque.

Other methods changes the CoM of the spacecraft. This is done by changing the mass distribution of the spacecraft. Concepts like adding moving mass ballasts to the sail booms[6], a 2-DOF control boom[7] or a 2-axis translation stage[8] between the satellite bus and the sail have been proposed to change the CoM. Changing the CoM can generate control torques in 2 axes. The different control methods for solar sails are described in greater described in by Fu et al[9].

A large number of the attitude control methods described above do not scale well with an increase in solar sail area. The 2-DOF control boom mass needs to increase if the sail size increases. The control vanes actuator does scale with an increase in sail size but with additional mechanics required at the corners of the sail does mean a larger $\lambda$ in comparison with the change in reflective properties method that IKAROS used.

### 3. Tri-Spin Solar Sail Satellite with Gyro Control

The CMG controlled tri-spin solar sailing satellite aims to provide an improved maneuverable platform for a deployed solar sail with a scalable attitude control strategy.

The tri-spin solar sail[10-11] is based on the need of improving the attitude maneuverability of a rotating solar sail. A rotating solar sail has many advantages above a stabilized sail. The main advantage is the constant centrifugal force, which keeps the sail stiff and reduces the requirement semi-rigid booms. This reduced specification on the booms will greatly reduce the mass and the deployment complexities. The major drawbacks of constantly spinning your spacecraft is the high angular momentum, which prevents fast attitude maneuvers and the complexity of operating body-mounted attitude actuators.

The conceptual satellite consists of a rotating sail section. In order to have a stabilized platform for payloads and attitude actuators the satellite bus is connected to the rotating sail structure by means of a rotating link like a motor. This creates a dual-spin satellite with the rotating sail an external momentum wheel. This external momentum wheel creates a large angular momentum bias, which greatly reduces the speed of attitude changes. In some applications, a momentum wheel within the satellite bus absorbs the angular momentum bias. This is not practical for large solar sails and thus a second rotating structure is attached to the satellite body. This rotating structure can simply deploy basic wire booms to produce a larger moment of inertia and thus reduce the rotation speed requirement to eliminate the angular momentum bias. Thus, the tri-spin solar sailing satellite consists of three sections (see Fig. 1) with the solar sail and the momentum countering system rotating relative to the stabilized satellite bus. The attitude of a tri-spin satellite can be controlled with attitude actuators mounted within the satellite bus.

Scalability of the attitude control method can be achieved by placing each of the rotating structures on a rotation stage. This creates a dual-gimbal CMG configuration as seen in Fig. 2. The main advantage of CMGs is that small changes in gimbal angles can produce very large control torques. The CMG controlled tri-spin has an attitude actuator, which scales with the moment of inertia and speed of the rotating structures, but does add mechanical complexity to the overall satellite design.

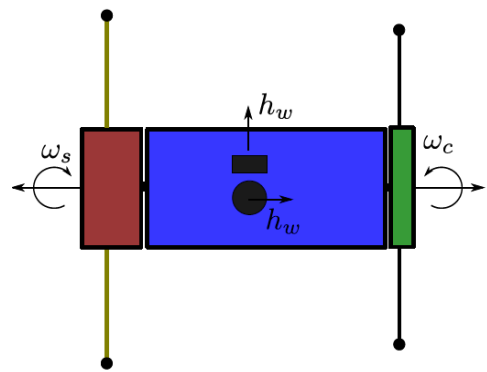

**Fig. 1.** The basic tri-spin solar satellite concept with the rotating sail in red, the satellite bus in blue and the moment countering system in green.

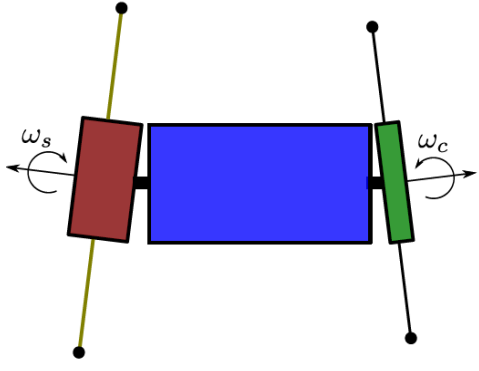

Fig. 2. The CMG controlled tri-spin solar satellite concept with the rotating sail and momentum countering system mounted on rotation stages.

## 4. Attitude Control with Gyro-Control Tri-Spin

The gimbal angles of the sail system and MCS is used to control the angular momentum and change in angular momentum of the satellite system. A steering law is required to determine the required changes in gimbal angles to generate a certain torque reference. A standard attitude controller such as a quaternion-feedback controller generates this torque reference. Fig. 3 illustrates the control system with the standard attitude controller and steering law.

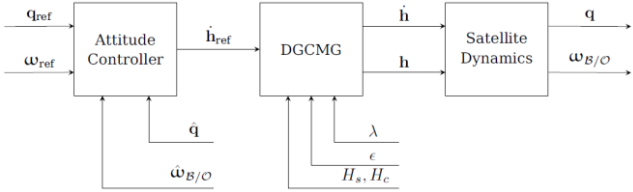

Fig. 3. Attitude control system layout for a Gyro-Control Tri-Spin satellite.

### 4.1. Satellite Attitude Dynamics

The attitude of the satellite is described with regards to two major reference frames, namely the inertial and body frames. The body frame is attached to the satellite and the attitude of the satellite is described by the transformation between the body frame and the inertial frame. This transformation from the body reference frame to the inertial frame $A_B^I$ is described by either Euler angles ($\varphi$, $\theta$, or $\psi$) or by a quaternion set ($\mathbf{q} = [q_1 \ q_2 \ q_3 \ q_4]^T$).

The standard Newton-Euler equation describing the attitude dynamics of a rigid satellite containing momentum exchange devices is:
$$\mathbf{N} = \mathbf{I}\dot{\boldsymbol{\omega}}_{B/I} + \boldsymbol{\omega}_{B/I} \times (\mathbf{I}\boldsymbol{\omega}_{B/I} + \mathbf{h}) + \dot{\mathbf{h}} \quad (3.)$$
where $\mathbf{N}$ is external torques on the satellite, $\mathbf{I}$ the moment of inertia of the entire satellite, $\mathbf{h}$ the angular momentum from internal sources and $\boldsymbol{\omega}_{B/I}$ the inertial referenced angular rates of the satellite. This equation can be extended to describe the dynamics of a satellite containing non-rigid elements
$$\mathbf{N} = \mathbf{I}\dot{\boldsymbol{\omega}}_{B/I} + \dot{\mathbf{I}}\boldsymbol{\omega}_{B/I} + \boldsymbol{\omega}_{B/I} \times (\mathbf{I}\boldsymbol{\omega}_{B/I} + \mathbf{h}) + \dot{\mathbf{h}} \quad (4.)$$
The effect of the non-rigid elements on the rotational dynamics of the entire spacecraft is describe with the addition of the time derivative of the moment of inertia $\dot{\mathbf{I}}$. Additional non-rigid dynamic models can describe the effects of the spacecraft's rotational dynamics on the relative position and movement of the non-rigid elements (see Fig. 4).

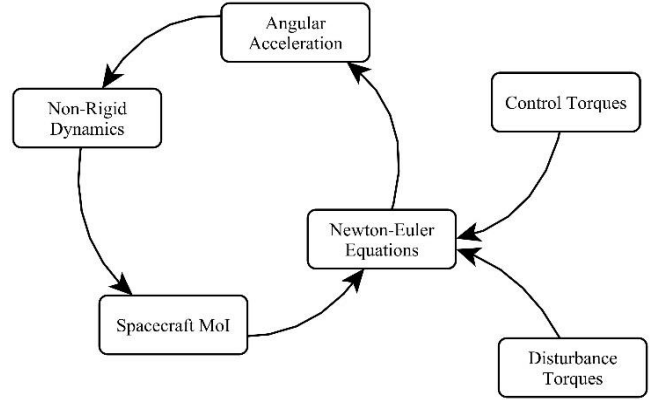

Fig. 4. Cross-coupling dynamics of a non-rigid satellite.

The tri-spin satellite contains three parts: the satellite bus, rotating sail and the rotating momentum counter system (MCS). The moment of inertia of the entire satellite is the sum of these parts
$$\mathbf{I} = \mathbf{I}_B + \mathbf{I}_S + \mathbf{I}_C \quad (5.)$$
with $\mathbf{I}_B$ the moment of inertia of central body, $\mathbf{I}_S$ the spinning deployed sail and $\mathbf{I}_C$ the spinning momentum counter system. The sources of internal angular momentum is
$$\mathbf{h} = \mathbf{I}_S\boldsymbol{\omega}_{S/B} + \mathbf{I}_C\boldsymbol{\omega}_{C/B} \quad (6.)$$
The angular rate of the sail $\boldsymbol{\omega}_{S/B}$ and the momentum counter system $\boldsymbol{\omega}_{C/B}$ relative to the satellite body determines the direction of the angular momentum. To maximize maneuverability of the satellite the angular momentum bias should be low, thus
$$\mathbf{h} = \mathbf{I}_S\boldsymbol{\omega}_{S/B} + \mathbf{I}_C\boldsymbol{\omega}_{C/B} \approx 0 \quad (7.)$$
A similar system is created by the MicroMAS satellite[12]. The MicroMAS is a dual-spin CubeSat with a payload rotating relative to the satellite body. It uses an internal momentum wheel to reduce the angular momentum bias created by the spinning payload. In this case, the payload's angular momentum contribution is small and it is practical to make use of an internal momentum wheel. In the case of a rotating solar sail, the angular momentum is too large to be countered by means of an internal momentum wheel. The momentum counter system is used to create a scalable method to reduce the angular momentum bias.

### 4.2. Attitude Controller

The two major limiting factors regarding the speed at which a solar sail satellite can change its attitude is the structural dynamics of the sail and the torque capabilities of its attitude actuators. Angular accelerations of the satellite body will result in oscillations in the sail surface or support structure. These oscillations of the non-rigid elements change the mass distribution of the satellite and therefore its moment of inertia. The change in moment of inertia will change the rotational dynamics of the satellite. This can clearly be seen in the satellite attitude dynamics described in Eq. (4).

When performing slow attitude maneuvers the bandwidth of the attitude controller is much slower than the structural dynamics of the satellite and smaller oscillations are induced in the non-rigid elements of the sail. A rotating sail produce centrifugal force, which increases the internal force of the sail material and other non-rigid elements and makes it more stiff. This internal force reduces the amplitude of the oscillations that occurs during attitude maneuvers. The attitude control should perform highly damped attitude maneuvers. Highly damped maneuvers reduce the duration of angular accelerations and therefore reduce the size of oscillations that occur.

A standard quaternion feedback controller[13] can be used to determine the control signals required to change the attitude described by a certain quaternion vector reference. The proportional and derivative control law for the required control signal is

$$\boldsymbol{u} = \boldsymbol{K}_d \boldsymbol{\omega}_{B/I} + \boldsymbol{K}_q \boldsymbol{q}_e \quad (3.)$$

with $\boldsymbol{\omega}_{B/I}$ the body rate and $\boldsymbol{q}_e$ the quaternion error which describes the rotation required from the current attitude to the required attitude. The values within the $\boldsymbol{K}_d$ and $\boldsymbol{K}_q$ matrices controls the settling time and damping of the attitude maneuver. The input to the attitude actuator, taking into account the current angular momentum of the spacecraft, becomes

$$\dot{\boldsymbol{h}} = -\boldsymbol{\omega}_{B/I} \times (\boldsymbol{I}\boldsymbol{\omega}_{B/I} + \boldsymbol{h}) - \boldsymbol{u} \quad (8.)$$

This control torque can be generated by a control moment gyro.

### 4.3. Tri-spin CMG control

The angular momentum of the rotating sail relative to the satellite body frame assuming the sail is rigid is described by

$$\boldsymbol{h}_S = \boldsymbol{I}_S \boldsymbol{\omega}_{S/B} = \boldsymbol{A}_S^B [0 \; H_S \; 0]^T \quad (9.)$$

with $\boldsymbol{A}_S^B$ the transformation matrix between the sail reference frame and the satellite body reference frame and $H_S$ is the angular momentum of the sail. The transformation matrix results the angular momentum of the sail to become

$$\boldsymbol{h}_S = \begin{bmatrix} -H_S C\epsilon_S S\lambda_S \\ H_S C\epsilon_S C\lambda_S \\ -H_S S\epsilon_S \end{bmatrix} \quad (10.)$$

with $C$ = cosine function and $S$ = sine function. The angles of the 2-DOF gimbal are described by $\lambda_S$ and $\epsilon_S$. Similarly the angular momentum of the MCS is defined as $\boldsymbol{h}_C$ so that if the gimbal angles are zero and $H_S \approx H_C$, then the total angular momentum would be $\boldsymbol{h} = \boldsymbol{h}_S + \boldsymbol{h}_C \approx \boldsymbol{0}$. The total angular momentum of the satellite can be described as

$$\boldsymbol{h} = \boldsymbol{A}(\boldsymbol{\delta}) \quad (11.)$$

where $\boldsymbol{\delta}$ designates the input vector of the DGCMG system. The torque is then defined as

$$\dot{\boldsymbol{h}} = \boldsymbol{A}(\boldsymbol{\delta})\dot{\boldsymbol{\delta}} \quad (12.)$$

The inverse of the equation above is used to determine the required gimbal inputs to generate the required torque. A scissored double-gimbal configuration is created when the gimbal angles of the sail system and the MCS are equal but opposite, thus $\epsilon = -\epsilon_S = \epsilon_C$ and $\lambda = -\lambda_S = \lambda_C$. The total angular momentum becomes

$$\boldsymbol{h} = \begin{bmatrix} -(H_S + H_C)C\epsilon S\lambda \\ (H_S - H_C)C\epsilon C\lambda \\ (H_S + H_C)S\epsilon \end{bmatrix} \quad (13.)$$

And the derivative is

$$\dot{\boldsymbol{h}} = \begin{bmatrix} -(\dot{H}_S + \dot{H}_C)C\epsilon S\lambda + \dot{\epsilon}(H_S + H_C)S\epsilon S\lambda - \dot{\lambda}(H_S + H_C)C\epsilon C\lambda \\ (\dot{H}_S - \dot{H}_C)C\epsilon C\lambda - \dot{\epsilon}(H_S - H_C)S\epsilon C\lambda - \dot{\lambda}(H_S - H_C)C\epsilon S\lambda \\ (\dot{H}_S + \dot{H}_C)S\epsilon + \dot{\epsilon}(H_S + H_C)C\epsilon \end{bmatrix} \quad (14.)$$

The torque matrix assuming that the torque requirement of the rotating systems are shared equally between the rotating sail and the MCS ($\dot{H} = \dot{H}_C = -\dot{H}_S$) is

$$\dot{\boldsymbol{h}} = \begin{bmatrix} 0 & (H_S + H_C)S\epsilon S\lambda & -(H_S + H_C)C\epsilon C\lambda \\ -2C\epsilon C\lambda & -(H_S - H_C)S\epsilon C\lambda & -(H_S - H_C)C\epsilon S\lambda \\ 0 & (H_S + H_C)C\epsilon & 0 \end{bmatrix} \begin{bmatrix} \dot{H} \\ \dot{\epsilon} \\ \dot{\lambda} \end{bmatrix} \quad (15.)$$

Thus the input vector is determined by

$$\begin{bmatrix} \dot{H} \\ \dot{\epsilon} \\ \dot{\lambda} \end{bmatrix} = \boldsymbol{A}^{-1}\dot{\boldsymbol{h}} \quad (16.)$$

The control inputs from the steering law is the angular rate of gimbal angles and the change in combined angular momentum of the sail and MCS. The limiting factors of the CMG system is the torque capabilities of the driving motors which controls the combined angular momentum of the sail and MCS and secondly the rate at which the gimbals angles can be changed. The required gimbal angle changes can be reduced by increasing the angular rate of the deployable structures but will increase mechanical wear. The torque due to gimbal angle changes can scale with an increase in MoI of the deployed structures. This CMG system can produce large torques perpendicular to the sail normal axis with the torque in the sail normal axis limited to the torque capabilities of the driving motor.

### 5. Simulation

A simulation is constructed to illustrate the attitude response the gyro-control tri-spin solar sail satellite achieves. The simulation is performed on a small solar sailing satellite containing the major parameters defined in Table 2. These values describes a 6U sized CubeSat satellite with an 80m$^2$ deployed sail.

Table 2. Simulation parameters for tri-spin solar sailing satellite

| Parameter | Value |
| --- | --- |
| Sail area | 80m2 |
| Sail boom length | 6.5m |
| MCS boom length | 3m |
| Sail mass | 0.384kg |
| Sail tip mass | 50g |
| MCS tip mass | 100g |
| Satellite bus | 8kg |

The simulation places the satellite in an orbit and attitude maneuvers are performed to certain attitude references relative to the inertial frame. Sensor models are ignored at this stage to isolate the response of the attitude controller and not be effected by estimator dynamics. Two different cases are investigated with the first investigated the effects of a perfectly known system. The second set of results investigate the effects of certain inconsistencies to the output. These errors includes the effect of momentum bias and some control inaccuracies. All the scenarios use the same quaternion feedback controller producing a highly damped system and a settling time of about 5 minutes for all attitude maneuvers.

### 5.1. Attitude Maneuvers

The performance of a gyro-controlled tri-spin satellite with perfect control is shown in Fig. 5 to Fig. 7. The simulation assumes a satellite with no momentum bias ($\mathbf{h} = \mathbf{h}_S + \mathbf{h}_C = \mathbf{0}$), the gimbal achieves the requested control references and the satellite model is completely known. The scenario is repeated with the sail spinning at 0.1rev/s and 1rpm.

The satellite is capable of achieving all the attitude references as seen in Fig. 5, with the response remaining unchanged with a change in sail angular rate. Small coupling between the axes are visible on the step responses due to the coupled nature of the steering law. Fig. 6 and Fig. 7 highlight the trade-off between control signal and angular momentum with much larger gimbal angles required for the 1rpm case.

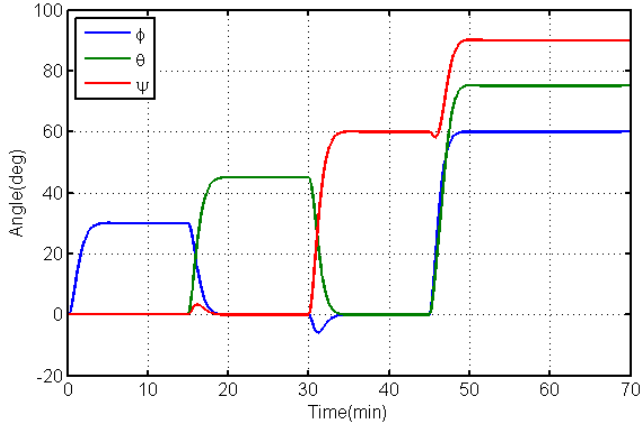

**Fig. 5. Step-responses to a number of different attitude references.**

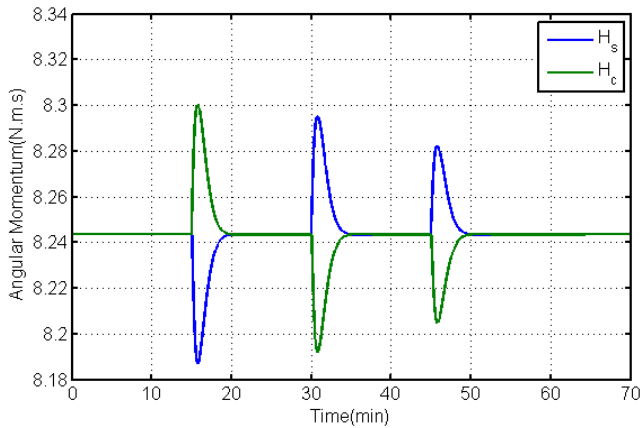

**Fig. 6. Angular momentum of sail and MCS for sail with angular rate 0.1rev/s.**

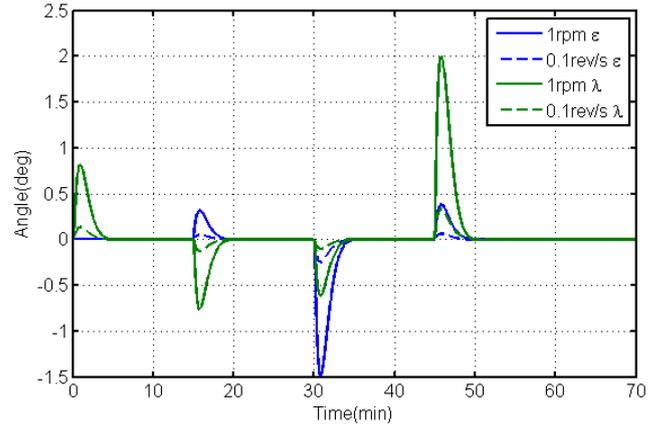

**Fig. 7. Gimbal angles for sail with angular rate of 0.1rev/s and 1rpm.**

### 5.2. Robustness Investigation

Different control errors are introduced in the system with the sail spinning at 1rev/s to investigate the effect on the response. Fig. 8 shows the response when an unknown angular momentum bias (negative to positive 5% of angular momentum of sail) is changed, thus $\mathbf{h} = \mathbf{h}_S + \mathbf{h}_C \neq \mathbf{0}$. The angular momentum bias results in larger gyroscopic disturbance torques, which the satellite absorbs with additional transient response.

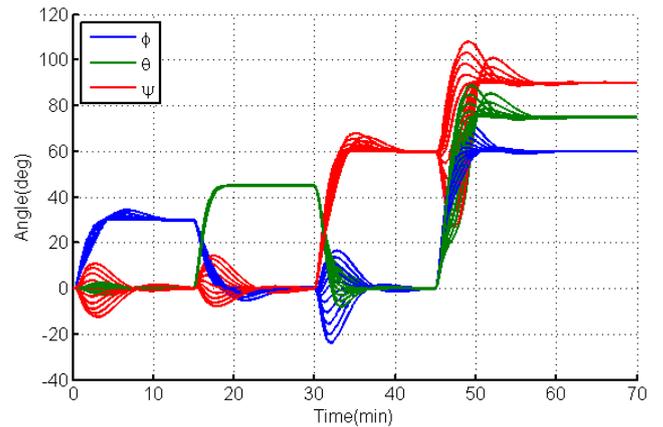

**Fig. 8. Step response with an angular momentum bias error.**

The scenario is investigated with an initial bias on the $\epsilon$ angle (from -1deg to 1deg) with the attitude response shown in Fig. 9.

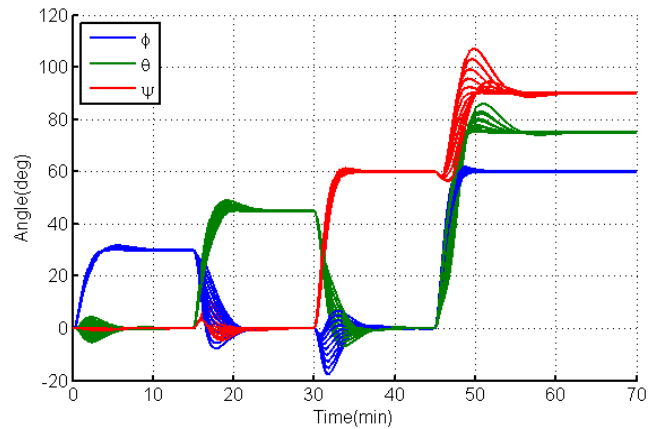

**Fig. 9. Step response with a moment of inertia error.**

The simulation is repeated with an error induced in the gimbal angles resulting in $-\epsilon_S \neq \epsilon_C$. This error ranges from -1deg to +1deg. The angle offset results in the angular momentum of the sail and the MCS not to cancel out and thus have a similar output than that of the angular bias case, see Fig. 10.

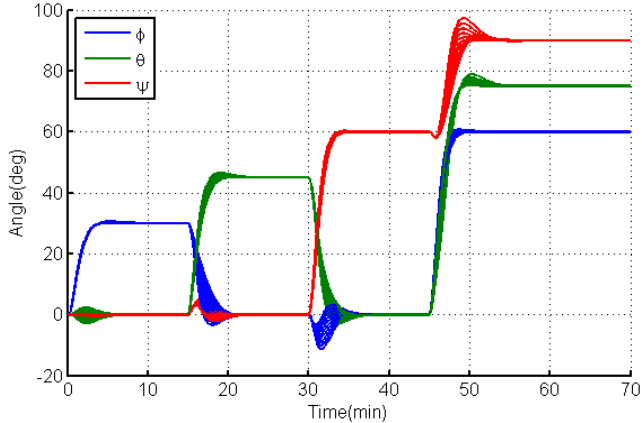

**Fig. 10.** Step response with gimbal angle error.

The robust analysis shows that the attitude maneuvers can be obtained with a number of errors. The effect of all these errors are determined by the angular momentum of the sail and MCS. If the angular rate of the sail is reduced, the effect of these errors will also be lowered. Reducing the angular rate of the sail does influence the centrifugal force within the sail and may be more affected by the angular accelerations of the spacecraft.

## 6. Conclusion

Future solar sailing satellites will be required to fulfill scientific measurements, which will need accurate and high performance attitude maneuvers. Current attitude control methods can control sailing satellites but do not scale well with an increase in sail size.

A scalable attitude control method, which can produce the torque required to complete fast attitude maneuvers successfully, were presented. The gyro-control uses a tri-spin satellite with a spinning sail system and momentum countering system each mounted on their respective 2-DOF gimbal. By changing the gimbal angles by applying the derived steering law the required torque can be generated. Simulations show that such a satellite can complete large attitude changes successfully. The responses were investigated with the addition of a number of error cases. Clearly, the attitude responses are greatly influenced by angular momentum bias, which is expected with large rotating structures. The angular momentum of the spacecraft needs to be well managed for this system to work optimally.

Although this conceptual attitude control system introduces additional mechanical complexities, it does provide a scalable attitude actuator solution for a spinning solar sail satellite.